# Pseudo Random Coins Show More Heads Than Tails

Heiko Bauke[1,*] and Stephan Mertens[1,2,†]

[1] *Institut f"ur Theoretische Physik, Otto-von-Guericke-Universit"at, Postfach 4120, 39016 Magdeburg, Germany*
[2] *The Abdus Salam International Centre for Theoretical Physics, St. Costiera 11, 34100 Trieste, Italy*

Tossing a coin is the most elementary Monte Carlo experiment. In a computer the coin is replaced by a pseudo random number generator. It can be shown analytically and by exact enumerations that popular random number generators are not capable of imitating a fair coin: pseudo random coins show more "heads" than "tails". This bias explains the empirically observed failure of some random number generators in random walk experiments. It can be traced down to the special role of the value zero in the algebra of finite fields.

## 1. MANUFACTURING RANDOMNESS

"After 40 years of development, one might think that the making of random numbers would be a mature and trouble-free technology, but it seems the creation of unpredictability is ever unpredictable." These words, written ten years ago by Brian Hayes [9], allude to the "Ferrenberg affair": In 1992, Ferrenberg, Landau and Wong [4] had shown that a well established family of pseudo random number generators produces wrong results in Monte Carlo simulations based on the Wolff algorithm. Since then, deficiencies in pseudo random number generators have been detected in various other simulations, like simulations with the Swendsen-Wang algorithm [1], $3d$ self avoiding random walks [7], the Metropolis algorithm on the Blume-Capel model [17] and $2d$ random walks [18, 19]. Certainly this list is incomplete (see the references in [19]), but the message is clear: Hayes' quote is still up-to-date. As long as the generation of random numbers is more an art than a science, the whole Monte Carlo method, indispensable tool in fields ranging from high-energy physics to economics, is based on shaky grounds. One crucial step on the way from art to science is a precise understanding of the mechanism behind the failures of pseudo random number generators in simulations. Explaining these failures is not as easy as detecting them, however. The 1992 "Ferrenberg affair" for example has been resolved only recently [16].

In this contribution we address the issue of why some pseudo random number generators lead to inconsistent results in random walk simulations [7, 18, 19]. A random walker in a lattice basically throws a coin to decide to go north or south, east or west, up or down. In $d$ dimensions, $d$ flips of a coin fix the next step of the walker. Adding reflecting or absorbing boundary conditions, a memory (self avoiding walks) or a site-dependent bias of the coin makes the walk more interesting but less transparent. Obviously a random number generator that is not capable of imitating a fair coin cannot work properly in more complex random walk experiments. Thus the basic question is: how well can a pseudo random number generator approximate a Bernoulli-$1/2$ process? This question is simple enough to be thoroughly analyzed, and the answer is somewhat surprising. The popular random number generators based on arithmetic modulo 2 are very bad approximations of a Bernoulli-$1/2$ process: Pseudo random coins produce more "heads" than "tails". This can be shown analytically. The problems arise because of the special role of the zero in the arithmetic of finite fields. Random number generators that avoid (or ignore) the zeros are much better representatives of a Bernoulli-$1/2$ process, in fact they are even better than real, physical coins [5].

This work was motivated by an exercise in Donald Knuth's seminal treatise on random number generators [12, exercise 3.3.2.31], and we start by repeating the experiment proposed in this exercise: We measure the probability that a run of length $w$ ($w$ odd) contains more "heads" than "tails" when produced by a recurrence relation in $\mathbb{Z}_2$. Surprisingly this probability is almost never even close to $1/2$! We explain how this probability can be calculated analytically using generating functions, and we demonstrate that the bias can be decreased, but not be removed by tinkering with the recurrence relation. It is a genuine feature of the arithmetic in $\mathbb{Z}_2$, or more precisely of the special properties of the zero element in finite fields. In $\mathbb{Z}_2$, the zero element is essential, but already in $\mathbb{Z}_3$ it can be circumvented to generate runs of "heads" and "tails" that are perfectly balanced. Last but not least we discuss how the empirically observed failure of random number generators in random walk experiments can be understood quantitatively.

## 2. PSEUDO RANDOM COINS

Almost all random number generators follow the same principle: they calculate a new random number from a subset of the previous numbers, i.e. they implement a

*E-mail:heiko.bauke@physik.uni-magdeburg.de
†E-mail:stephan.mertens@physik.uni-magdeburg.de

recurrence

$$x_k = f(x_{k-1}, x_{k-2}, \dots, x_{k-p})\,. \tag{1}$$

to generate a sequence $(x_k)$ of pseudo random numbers. For a practical algorithm the numbers $x_k$ are from a finite domain, and without loss of generality we can assume that this domain is a *finite field*. For simplicity we take this field to be $\mathbb{Z}_m$, the numbers $0, 1, \dots, m-1$ with multiplication and addition modulo $m$. $\mathbb{Z}_m$ is a field if and only if $m$ is prime. Now *any* periodic (or ultimately periodic) sequence over a finite field can be written as a linear recurrence [10], which in our case reads

$$x_k = a_1 x_{k-1} + a_2 x_{k-2} + \dots + a_p x_{k-p} \bmod m\,. \tag{2}$$

For a "random coin generator" it is natural to choose $m = 2$, with 0 denoting "head" and 1 denoting "tail". The coefficients $a_k$ then are either 0 or 1, and a particularly simple recurrence has only two feedback taps

$$x_k = x_{k-p} + x_{k-q} \bmod 2 \qquad k > p > q\,, \tag{3}$$

also known as linear feedback shift register (LFSR) sequence or $\mathrm{R}(p,q)$ generator. Note that addition modulo 2 is equivalent to the exclusive-or operation, hence (3) can be applied bitwise to multi-bit words to generate a stream of integers $x$. Since the single bits of these integers do not interact we will discuss the one-bit version $x_k \in \mathbb{Z}_2$ without too much loss of generality. Pseudo random number generators based on (3) have been introduced into the physics community in 1981 [11] as a very fast and reliable method, but they were proposed already in the 1950s [8]. The sequence (3) is periodic, but if $p$ and $q$ are chosen such that the so called feedback polynomial

$$x^p - x^q - 1 \tag{4}$$

is primitive modulo 2, the $\mathrm{R}(p,q)$ generator attains the maximum period $T = 2^p - 1$ [6, 10].

From a fair coin we expect to see 0 and 1 with equal probability $1/2$, doublets $(0,0)$, $(0,1)$, $(1,0)$ and $(1,1)$ should each appear with probability $1/4$ and so on. In general, each possible tuple $(x_1, \dots, x_w)$ of size $w$ should appear with probability $2^{-w}$. For an LFSR sequence with primitive feedback polynomial one can *prove* that each possible tuple of size $w$ occurs $2^{p-w}$ times per period for $w \le p$, except the all zero tuple which occurs $2^{p-w} - 1$ times. This is very close to the statistics of tuples in true random sequences, and this is why LFSR sequences with primitive feedback polynomial qualify as *pseudo noise sequences* [6, 10]. In terms of Compagner's ensemble theory[2, 3], tuples of size equal or less than $p$ drawn from a pseudo noise sequence are indistinguishable from true random tuples.

For tuples larger than the register length $p$ the equidistribution can no longer hold, however. The first $p$ bits of the sequence (3) determine the whole sequence, hence we have at most $2^p$ different tuples of any size in our sequence, i.e. only an exponential small fraction of all possible $w$-tuples actually occurs in LFSR sequences if $w > p$. An obvious example of a missing tuple is a run of $p+1$ successive ones. Such impossible tuples are inevitable as long as the random number generator is a finite state automaton, but it is not obvious whether these missing tuples do affect a simulation.

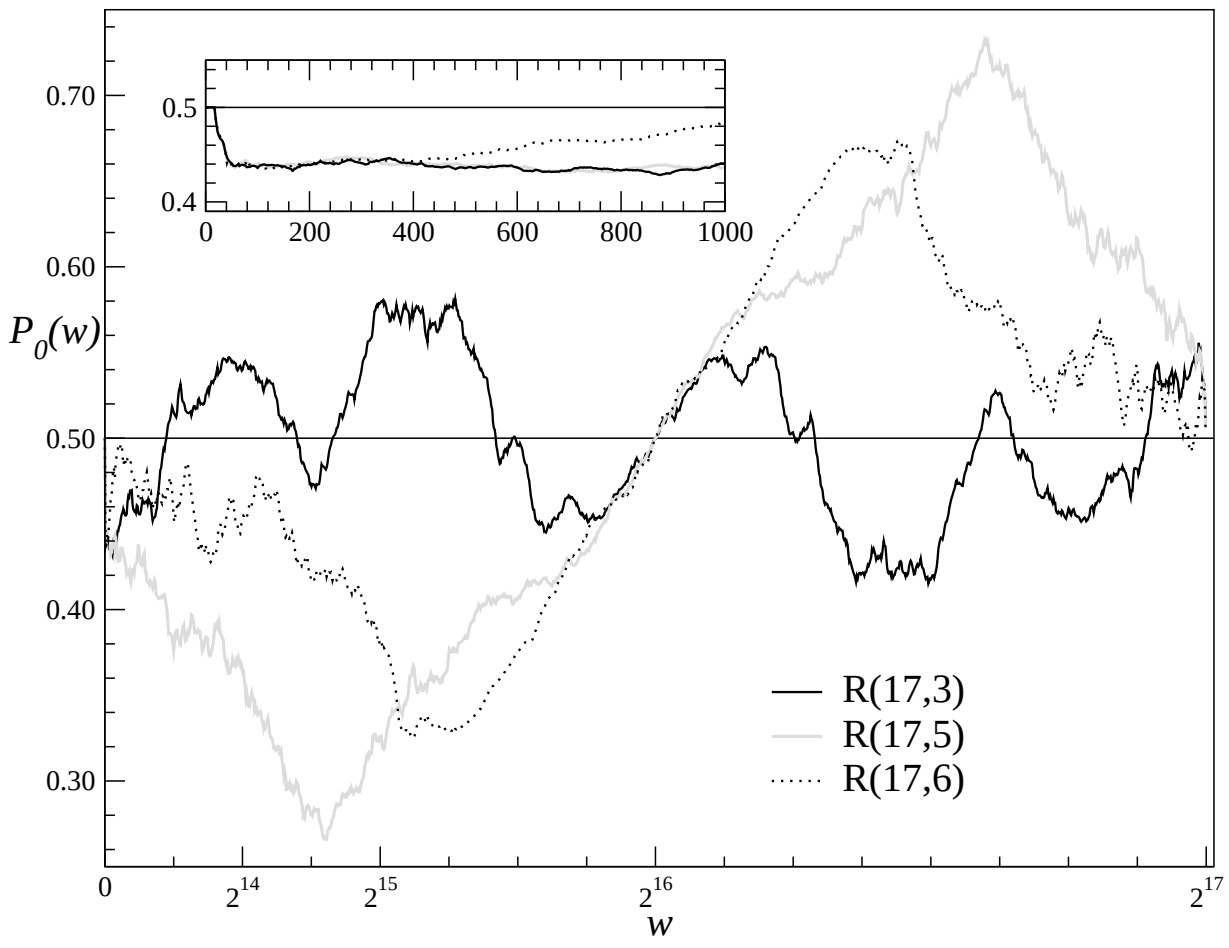

FIG. 1: Probability that a $w$-tuple ($w$ odd) of successive bits contains more zeros than ones for different pseudo random number generators $\mathrm{R}(p,q)$ with register length $p = 17$.

Most applications rely on more global properties of the bit sequence and are not sensitive to the absence or presence of specific patterns of bits. The classical example is the random walk, where the random bits are used to decide which direction to go in the next step. For walks in one dimension it is only the total number of ones vs. zeros in $w$-tuples that determine the position of the walker after $w$ steps. So let us look at the most coarse grained measure in our coin experiment: the probability $P_0(w)$ that after an odd number $w$ of flips of a coin we have more "heads" (1) than "tails" (0). $P_0(w) = 1/2$ for a fair coin, and for pseudo noise sequences we have

$$P_0(w) = \frac{2^{p-1} - 1}{2^p - 1} = \frac{1}{2} - \mathcal{O}(2^{-p}) \qquad \text{for } w \le p\,. \tag{5}$$

The deviations from $1/2$ are due to the missing all zero tuple, but they can safely be neglected for the values of $p$ that are used in practical random number generators ($p \ge 250$). The question is whether $P_0(w)$ stays near the $1/2$ for larger values of $w$ or not.

Fig. 1 shows $P_0(w)$ for $w$ ranging over the whole period of pseudo noise sequences with $p = 17$ and $q \in \{3, 5, 6\}$. The deviations from $1/2$ are striking. Tuples of a pseudo noise sequences $\mathrm{R}(17,5)$ of sizes around 26 000 (about one fifth of the period), for example, have a probability of less than 0.27 to contain more zeros than ones. One might object that these large deviations only appear for tuples that span substantial fractions of the whole period and that this regime is never sampled in simulations that use large values of $p$. This might hold for the extreme deviations, but $P_0(w)$ differs significantly from $1/2$ as soon as $w > p$, as can be seen from the inset in Fig. 1. Below we will calculate $P_0(w)$ for small $w$ and arbitrary $p$ to demonstrate that this disturbing bias is present for all values of $p$.

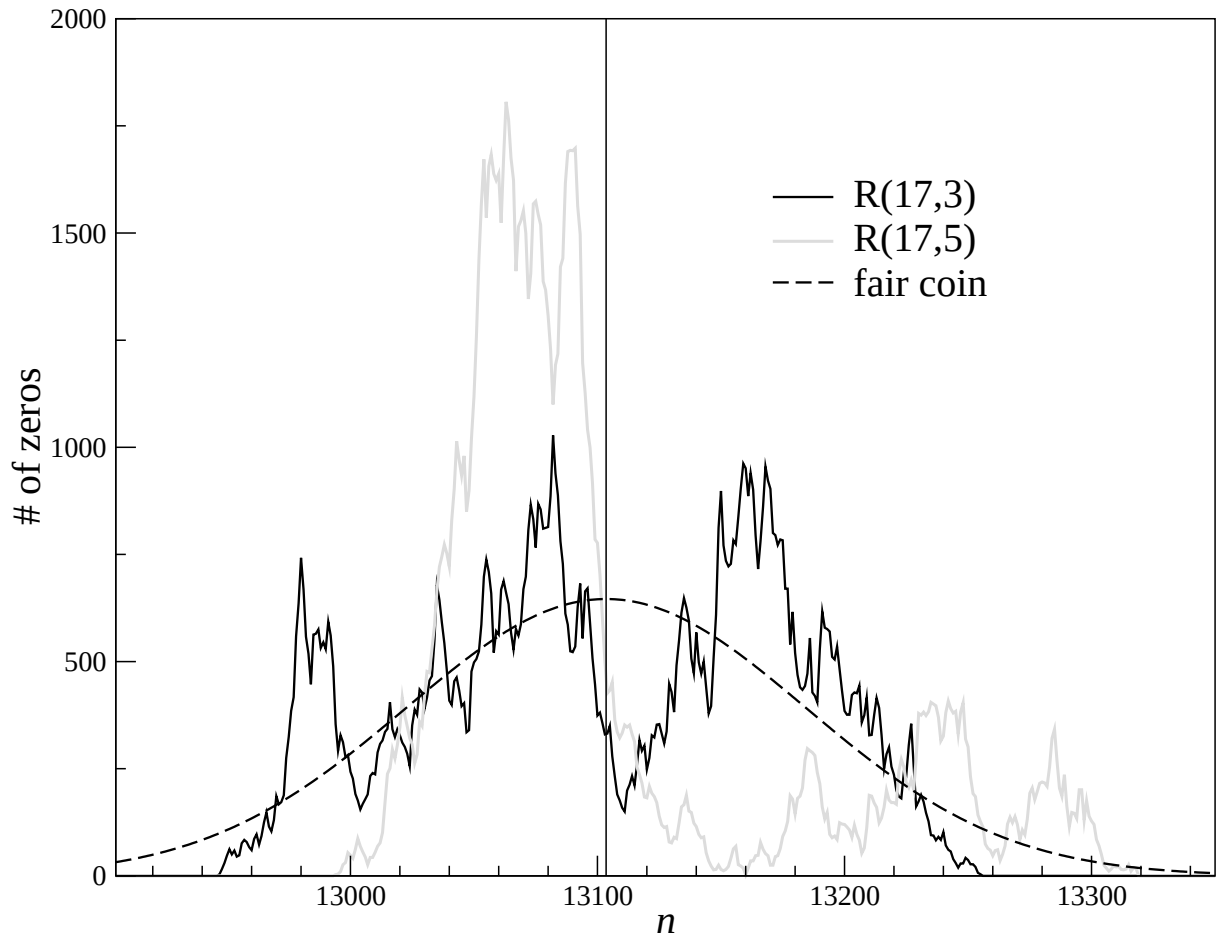

FIG. 2: Number of tuples of size $w = 26207$ that contain $n$ zeros as generated by R(17, 3) and R(17, 5). The dashed line is the theoretical outcome of a Bernoulli-$1/2$ process (fair coin).

The bias in $P_0(w)$ can be attributed to the *clustering of zeros*: Eq. (3) maps a block of zeros somewhere in the tuple to another block of zeros in the tuple. Hence we expect a majority of zeros in a tuple to be a *distinct* majority. This can be seen from Fig. 2, where we have displayed the distribution of zeros in tuples of size $w = 26\,207$. For these tuple size, the output of R(17, 5) is extremely biased ($P_0(w) = 0.265$), and Fig. 2 indeed shows a concentration of tuples with a number of zeros above $w/2$. For R(17, 3) the bias is much smaller ($P_0(w) = 0.482$) and the tuples with a majority of zeros concentrate closer to $w/2$. Both distributions differ significantly from the binomial distribution of a Bernoulli-$1/2$ process. Note that the average number of zeros in tuples of size $w$ is $w/2$ for all values of $w$.

A palpable feature of the curves $P_0(w)$ is their symmetry. Let $T = 2^p - 1$ denote the period of the pseudo noise sequence. Then $P_0(w)$ seems to be point symmetric around $T/2$, i.e. the curves look like $P_0(T-w) = 1-P_0(w)$ [20]. An inspection of the raw data reveals that the symmetry is not exact, however. Consider a tuple of odd size $w$ that contains more ones than zeros (this occurs with probability $1-P_0(w)$). Then the complementary tuple of even size $T - w$ cannot have more ones than zeros. It either has less ones than zeros (with probability $P_0(T-w)$) or the same number of ones and zeros. The latter means that the $w$-tuple must contain exactly $(w-1)/2$ zeros. We get

$$P_0(T - w) = 1 - P_0(w) - b(w) \tag{6}$$

where $b(w)$ denotes the probability that a tuple of odd size $w$ is balanced, i.e. that it contains exactly $(w-1)/2$ zeros. Eq. (6) is an exact equation but not exactly what we want, since it relates odd-sized tuples and even-sized tuples whereas Fig. 1 shows $P_0(w)$ for odd $w$ only. So let us relate the $w$-tuple to the tuple of odd size $T - 1 - w$. Now there is one bit left that is either one or zero. If we assume that this bit is uncorrelated with the bits in the tuples, we get

$$P_0(T - 1 - w) \approx 1 - P_0(w) - \frac{b(w)}{2}\,. \tag{7}$$

Eq. (7) is only an approximation because it neglects the correlation of the spare bit with the rest of the sequence, yet it provides us with a qualitative understanding of the near-symmetry in $P_0(w)$. For $w < p$, Eq. (7) is exact and can be simplified to

$$P_0(T - 1 - w) = \frac{1}{2} - \frac{1}{2^{w+1}}\binom{w}{\frac{w-1}{2}} + \mathcal{O}(2^{-p})\,, \quad w < p\,. \tag{8}$$

Note that the symmetry in $P_0(w)$ has nothing to do with the LFSR method. It is a sole consequence of the periodicity and the balance of zeros and ones within one period.

## 3. CALCULATING THE BIAS

Linear feedback shift register sequences are simple enough to allow an exact calculation of $P_0(w)$ at least for small $w$ [12]. Let $p_1(w, n)$ denote the probability of having $n$ one-bits in a tuple of size $w$, and let $f_w(z)$ denote the generating function of $p_1(w, n)$, i.e.

$$f_w(z) = \sum_{n=0}^{w} p_1(w, n) z^n\,. \tag{9}$$

Once $f_w(z)$ is known, the probability $P_0(w)$ of having more zeros than ones can be calculated easily,

$$P_0(w) = \sum_{n=0}^{(w-1)/2} p_1(w, n) \qquad \text{for odd } w\,. \tag{10}$$

Ignoring the $\mathcal{O}(2^{-p})$ corrections we know that each bit in tuples of size $w < p$ has probability $1/2$ to be zero or one, hence

$$f_w(z) = \left(\frac{1+z}{2}\right)^w, \qquad w \leq p\,. \tag{11}$$

For $w > p$, triples of bits appear that are related by (3). This can be depicted by a 3-uniform hypergraph: the vertices of the hypergraph are the bits of the tuple and a hyperedge joins each triple of nodes that are related by (3). Fig. 3 shows the hypergraph for $(p, q) = (17, 6)$ and $w = 19$. To calculate the generating function of the whole graph we need to know the generating functions of its unconnected components (Fig. 4).

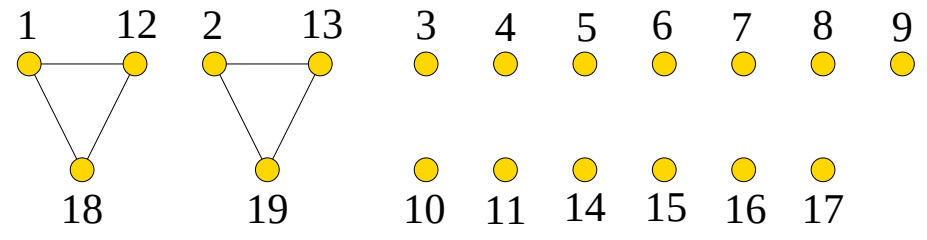

FIG. 3: Hypergraph representing tuples of size $w = 19$ in sequences generated by R(17, 6).

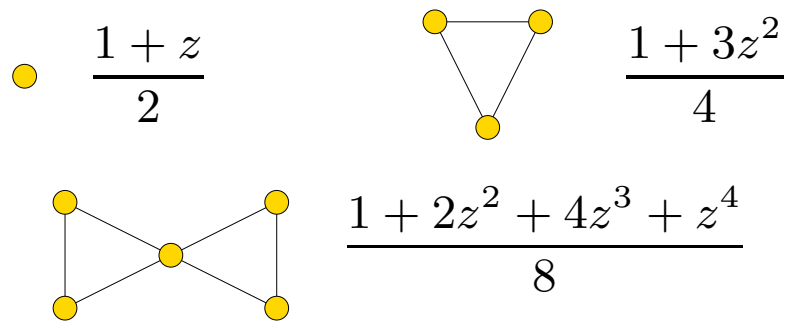

FIG. 4: Small components and their generating functions.

The generating function for the triangle reflects the fact that $x = y + z \bmod 2$ implies that either none (one configuration) or two (three configurations) bits equal one. For our example with $(p, q) = (17, 6)$ we obtain

$$f_{19}(z) = \left(\frac{1+z}{2}\right)^{13} \left(\frac{1+3z^2}{4}\right)^{2}. \tag{12}$$

This can easily expanded using a computer algebra system to get

$$P_0(19) = \frac{32\,053}{65\,536} \approx 0.4891\,. \tag{13}$$

As $w$ gets larger, more and more hyperedges (triangles) are added to the graph and the value of $P_0(w)$ decreases. Beyond a certain density of edges, triangles merge to form bow ties, see Fig. 5. The generating function for the bow tie type subgraph can easily be calculated (see Fig. 4) and the generating function that corresponds to Fig. 5 is

$$f_{25}(z) = \left(\frac{1+z}{2}\right)^{3} \left(\frac{1+3z^2}{4}\right)^{4} \left(\frac{1+2z^2+4z^3+z^4}{8}\right)^{2}. \tag{14}$$

Again we apply a computer algebra system to expand this polynomial and to sum up the coefficients to get

$$P_0(25) = \frac{15\,485}{32\,768} \approx 0.4726\,. \tag{15}$$

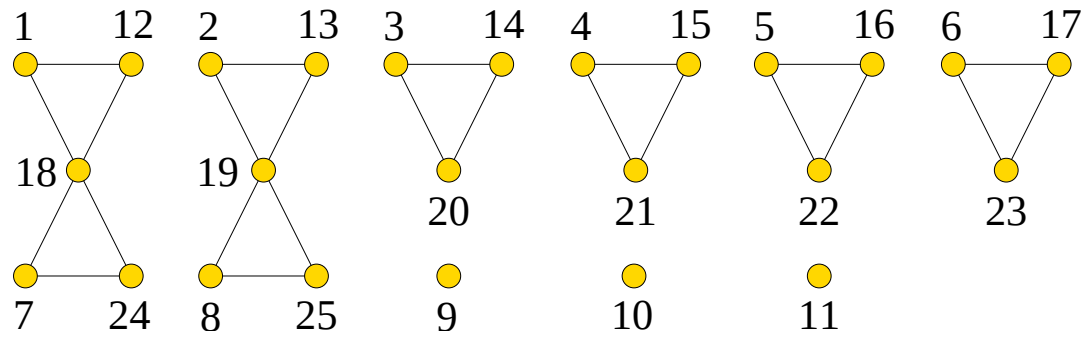

FIG. 5: Hypergraph representing tuples of size $w = 25$ in sequences generated by R(17, 6).

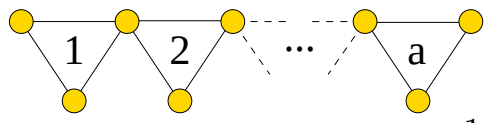

$$\begin{pmatrix} h(z) \\ g(z) \end{pmatrix} = \frac{1}{2^a} \begin{pmatrix} 1 & z \\ z^2 & z \end{pmatrix}^{a-1} \begin{pmatrix} 1 \\ z \end{pmatrix}$$

FIG. 6: The generating function $f(z) = h(z) + g(z)$ for a hyperpath of length $a$ can be calculated with $\mathcal{O}(\log a)$ matrix multiplications (using fast exponentiation).

If we increase $w$ further, more and more nodes are joined by hyperedges, new types of connected components appear and the calculation of the corresponding generating functions gets more complicated. Fig. 6 shows how to calculate the generating function for a path in the hypergraph. Beyond a certain value of $w$, the hypergraph is connected. The generating function can still be calculated exactly but we don't know a general method that needs less than $\mathcal{O}(2^p)$ operations. This limits the practical calculation of $P_0(w)$ to either small values $p$ (say $p < 40$) or to small values of $w$, where the hypergraph consists of disconnected, small components (see below).

We have seen how adding hyperedges to the graph decreases $P_0(w)$, but at some point this trend must reverse, last but not least because of Eq. (7). Complete graphs with $P_0(w) > 1/2$ are too big and entangled to be drawn here (see Fig. 8 for a partial graph of five vertices), but the mechanism that leads to an increase of $P_0(w)$ can be understood without seeing an example. Each hyperedge imposes more constraints on the variables in the graph, reducing the number of possible assignments. No set of constraints can ever rule out the all-zero assignment, however, hence zero-bits are favored in highly connected subgraphs.

So far our examples had small values of $p$, allowing us to enumerate the complete period of the pseudo noise sequence. Practical random number generators have large values of $p$, and simulations consume only a negligible part of the period. The general case can be analyzed easily as long as

$$w \le \min(p + q, 2p - q)\,, \tag{16}$$

since for these values of $w$ our hypergraph consists of isolated vertices and isolated triangles only. Setting $k = \max(w - p, 0)$ (the number of triangles) we get

$$f_w(z) = \left(\frac{1+z}{2}\right)^{p-2k} \left(\frac{1+3z^2}{4}\right)^{k} \tag{17}$$

and

$$p_1(n) = \frac{1}{2^p} \sum_{j=0}^{k} 3^j \binom{k}{j} \binom{p-2k}{n-2j}. \tag{18}$$

To calculate $P_0(w)$ we need to sum $p_1(n)$ from $n = 0$ to $n = (w-1)/2$. We haven't found a compact expression for this sum in general, but for $w = p + 1$ ($p$ even) and

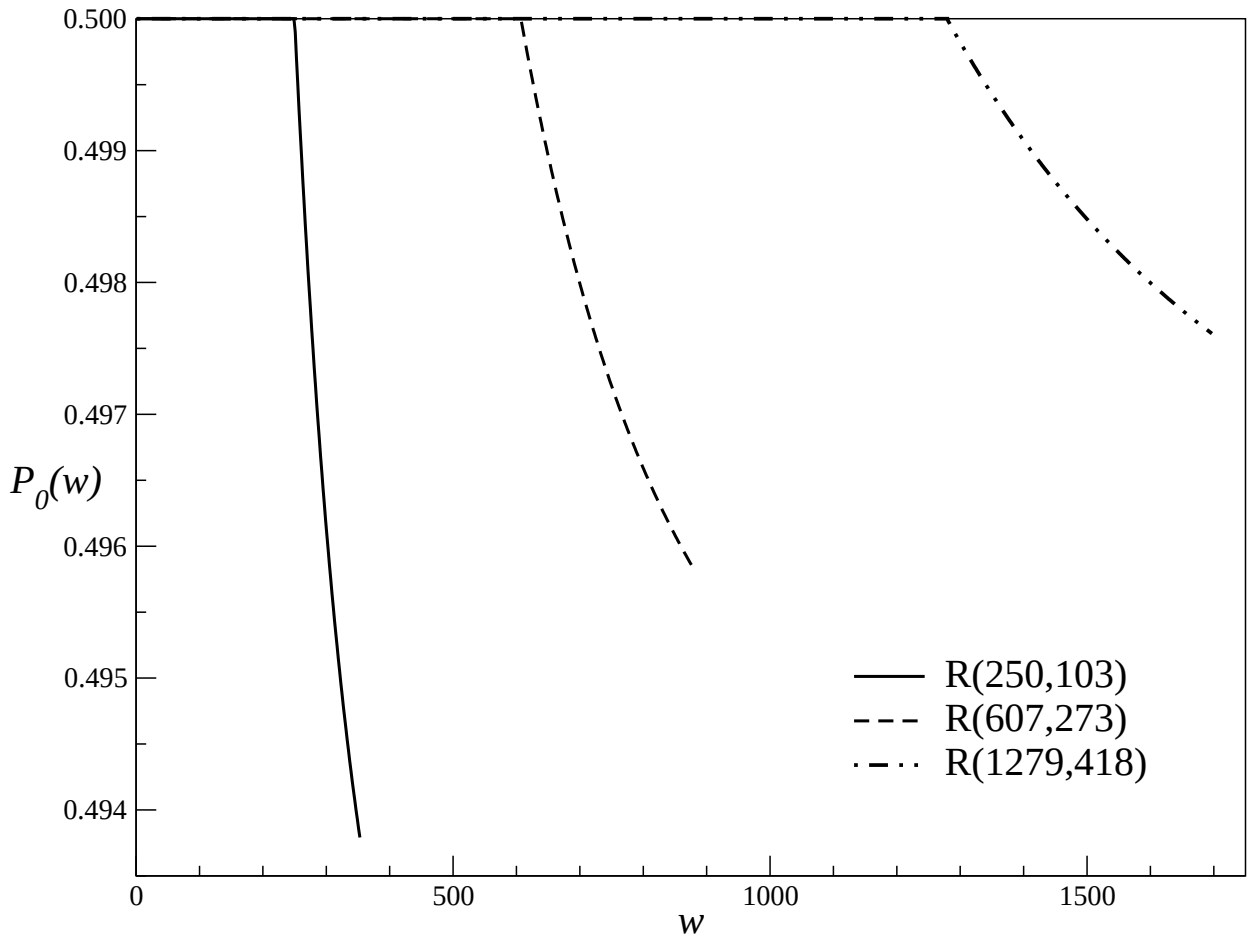

FIG. 7: Bias in tuples generated by "industrially sized" pseudo random number generators.

$w = p + 2$ ($p$ odd) the summation can be done:

$$\begin{aligned} P_0(p+1) &= \frac{1}{2} - \frac{1}{2^{p+1}(p-1)}\binom{p}{p/2} \\ &= \frac{1}{2} - \frac{1}{\sqrt{2\pi p^3}} + \mathcal{O}(p^{-5/2}) \end{aligned} \tag{19}$$

$$\begin{aligned} P_0(p+2) &= \frac{1}{2} - \frac{1}{2^p p}\binom{p}{(p+1)/2} \\ &= \frac{1}{2} - \frac{1}{\sqrt{\frac{\pi}{2}p^3}} + \mathcal{O}(p^{-5/2})\,. \end{aligned} \tag{20}$$

Fig. 7 shows $P_0(w)$ calculated by summing up $p_1(n)$ (18) numerically for values $(p, q)$ used in practical random number generators. Even for the small tuple sizes the bias is large enough to affect moderately precise simulations. The curves end at values of $w$ where (16) ceases to hold. For the generator $\mathrm{R}(9689, 471)$ discussed in [19], this value is $w = 10\,159$, and $P_0(10\,159) \approx 0.499\,817$.

## 4. MORE FEEDBACK

Pseudo noise sequences over $\mathbb{Z}_2$ starred in the "Ferrenberg affair" and they were caught to produce bad results in random walk simulations. On the other hand they do pass many statistical tests and they lead to extremely fast random number generators, so people tried to "improve" the quality of LFSR sequences, mainly based on empirical considerations. One proposal was to increase the number of feedback taps [19], i.e. to use

$$\begin{aligned} x_k &= x_{k-s} + x_{k-r} + x_{k-q} + x_{k-p} \bmod 2 \\ & \qquad k > p > q > r > s \end{aligned} \tag{21}$$

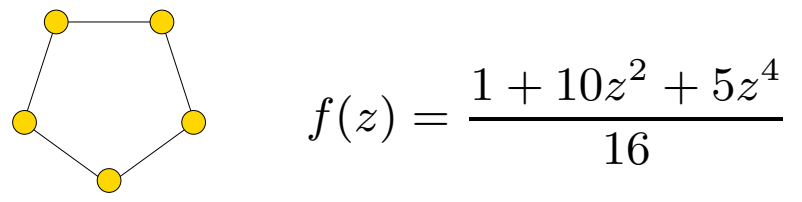

FIG. 8: Generating function for variables connected by recursion with $t = 4$ feedback taps.

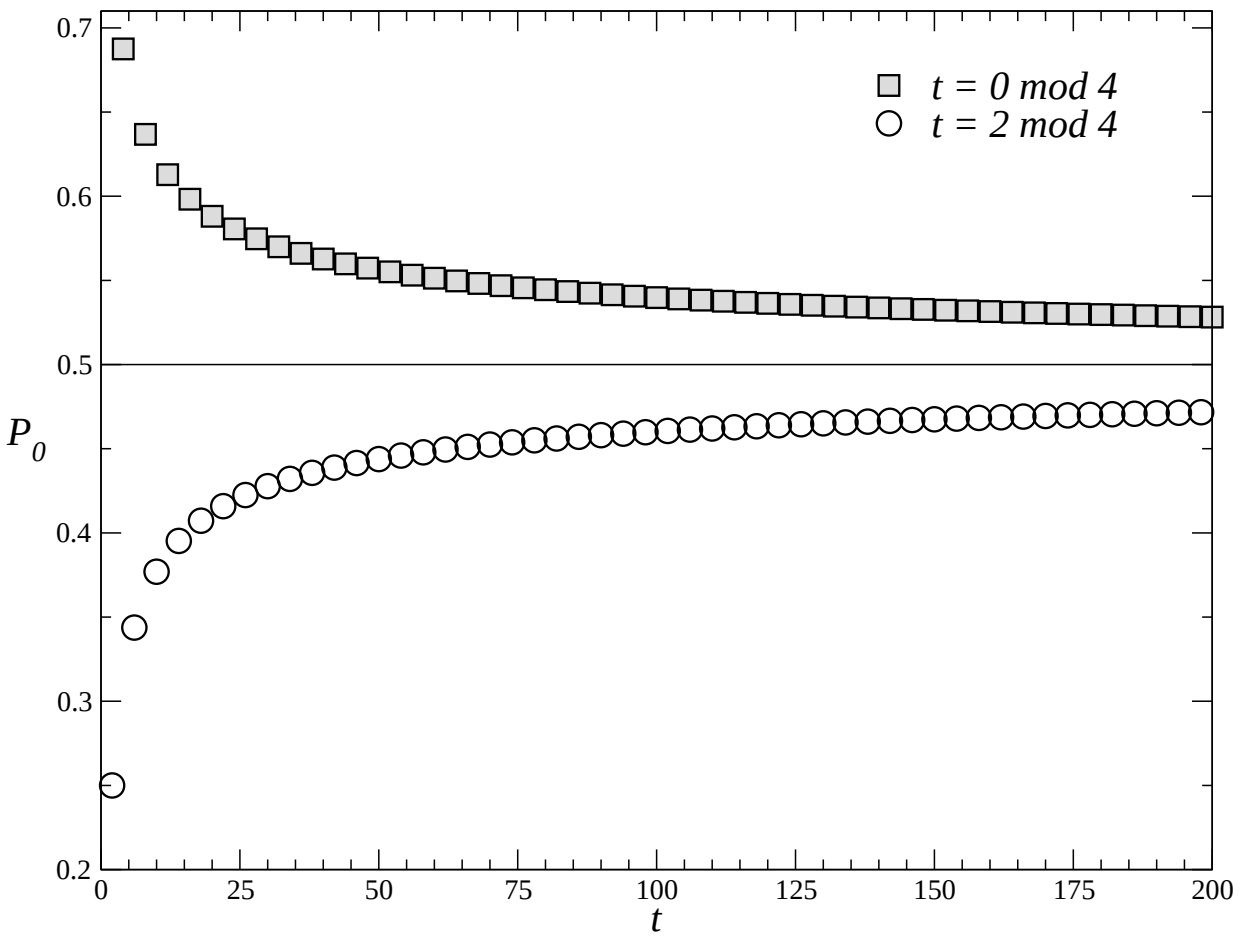

FIG. 9: Probability $P_0$ to have more zeros than ones in $t+1$ vertices connected by a single hyperedge. $t$ is the number of feedback taps.

with four feedback taps instead of two. If the corresponding feedback polynomial

$$x^p - x^q - x^r - x^s - 1 \tag{22}$$

is primitive modulo 2, the resulting sequence is again a pseudo noise sequence with period $T = 2^p - 1$. Note that the number of feedback taps must be even for a primitive polynomial to exist.

In fact it has been shown recently that increasing the number of feedback taps alleviates the failure of pseudo noise sequences in Monte Carlo simulations with the Wolff algorithm. The number of taps must be very large to reach the quality of other, much simpler generators, however [16]. The question is how the number of feedback taps affects the bias in our simple coin flipping experiment.

Due to the pseudo noise property we have $P_0(w) = 1/2 + \mathcal{O}(2^{-p})$ for $w \leq p$ independently of the number of feedback taps. For larger values of $w$, $P_0(w)$ depends on the number and the position of the feedback taps. If $t$ denotes the number of feedback taps, the resulting hypergraph is $(t+1)$-uniform, i.e. a single hyperedge connects $t+1$ variables. Fig. 8 shows the case $t = 4$ and its generating function. The generating function for $t+1$ variables connected by a hyperedge reads

$$f(z) = \frac{1}{2^t}\sum_{k=0}^{t/2}\binom{t+1}{2k} z^{2k}\,. \tag{23}$$

A hyperedge favors ones for $t \equiv 2 \mod 4$ and zeros for $t \equiv 0 \mod 4$, and the bias that is introduced with each hyperedge decreases with increasing $t$ (Fig. 9). If the $w$-tuple is small enough the graph consists of isolated points and isolated hyperedges. For these cases the generating function reads

$$f(z) = \left(\frac{1}{2^t}\sum_{k=0}^{t/2}\binom{t+1}{2k}z^{2k}\right)^{w-p}\left(\frac{1+z}{2}\right)^{p-t(w-p)}, \tag{24}$$

where $p$ is the largest feedback tap. Assuming $p$ even we get after some algebra

$$\begin{aligned} P_0(p+1) &= \frac{1}{2} + (-1)^{t/2}\frac{1}{2^{p+1}}\frac{\binom{p}{p/2}\binom{p/2}{t/2}}{\binom{p}{t}} \\ &= \frac{1}{2} + (-1)^{t/2}\frac{(t-1)!!}{\sqrt{2\pi p^{t+1}}} + \mathcal{O}(p^{-\frac{t+3}{2}}) \end{aligned} \tag{25}$$

and for odd $p$

$$\begin{aligned} P_0(p+2) &= \frac{1}{2} + (-1)^{t/2}\frac{1}{2^{p}}\frac{\binom{p}{(p+1)/2}\binom{(p+1)/2}{t/2}}{\binom{p+1}{t}} \\ &= \frac{1}{2} + (-1)^{t/2}\frac{(t-1)!!}{\sqrt{\frac{\pi}{2}p^{t+1}}} + \mathcal{O}(p^{-\frac{t+3}{2}}) \,. \end{aligned} \tag{26}$$

$n!! = n \cdot (n-2) \cdot \ldots$ denotes the double factorial function. In both cases the bias decreases like $\mathcal{O}(p^{-(t+1)/2})$, so we may conclude that increasing the number of feedback taps indeed alleviates the bias, at least for small values of $w$, where the hypergraph is sparsely connected. For larger values of $w$ the bias gets as large as in the case $t = 2$, however, even for dense feedback polynomials with $t \approx p/2$ (Fig. 10). Practical applications usually consume only tiny fractions of the period, hence they may well take advantage of the initial reduction of the bias for larger values of $t$. This explains the empirical observations of Ziff [19]. He proposed the four tap generator R(9689, 6988, 1586, 471), and for this generator (24) holds up to $w = 10\,159$. We get $P_0(10\,159) \approx 0.500\,000\,054$, a value that needs to be compared to $P_0(10\,159) \approx 0.499\,817$ of the two tap generator R(9689, 471). We observe the $\mathcal{O}(p^{-1})$ decrease of the bias predicted by (26) for $w = p + 2$. Increasing the number of taps apparently *suppresses* the bias, but Fig. 10 tells us that it does not truly *remove* it.

## 5. AVOID THE ZEROS

The notion of a pseudo noise sequence can be generalized to linear feedback shift register sequences over Galois fields like $\mathbb{Z}_m$ with $m$ being prime. The LFSR sequence

$$x_k = a_1 x_{k-1} + a_2 x_{k-2} + \ldots + a_p x_{k-p} \bmod m \tag{27}$$

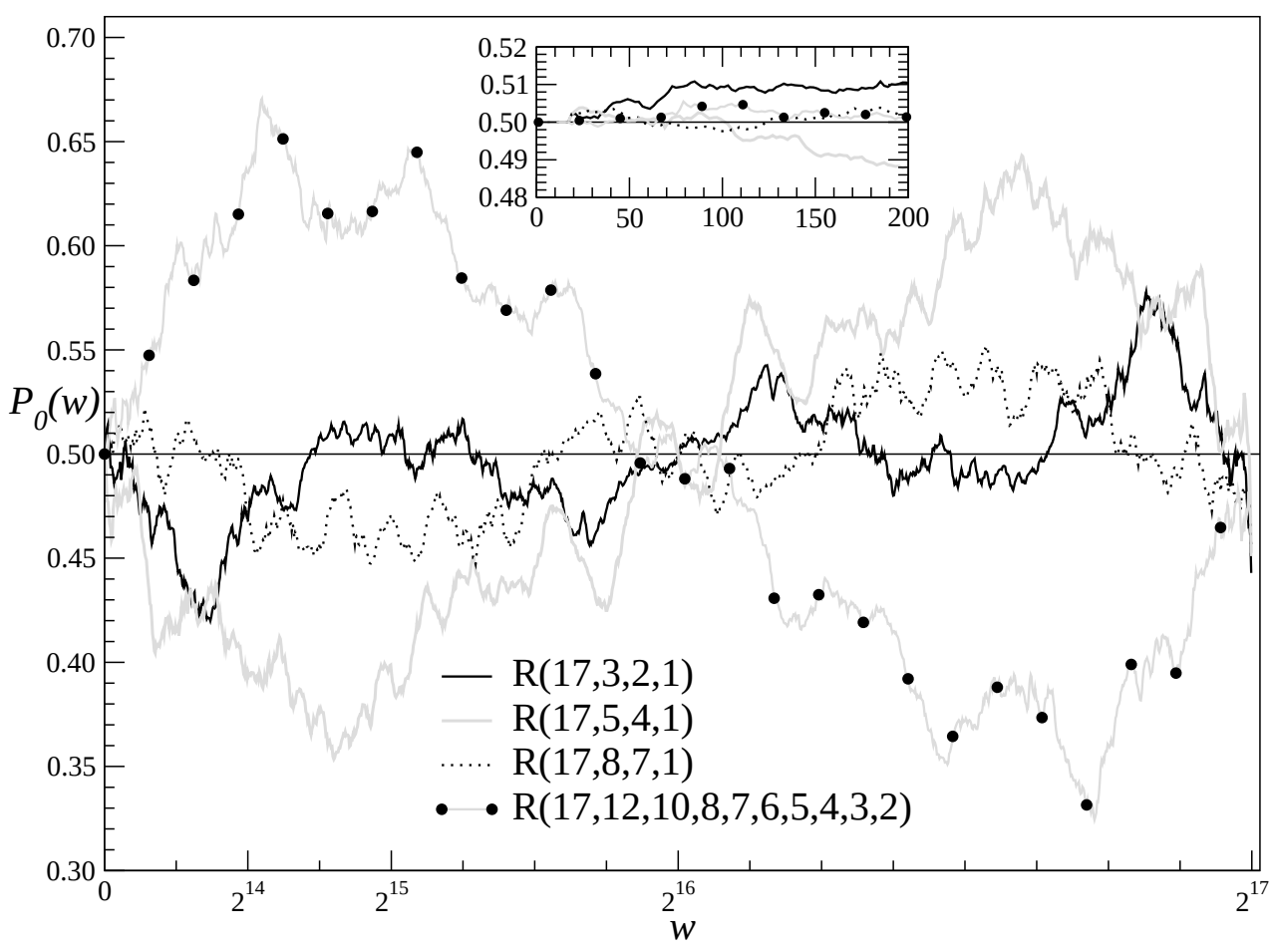

FIG. 10: Probability $P_0(w)$ that a $w$-tuple of successive bits of a pseudo noise sequence contains more zeros than ones for LFSRs with four and eight feedback taps.

with coefficients $a_i \in \mathbb{Z}_m$ attains the maximum period $T = m^p - 1$ if and only if the feedback polynomial

$$x^p - a_1 x^{p-1} - \ldots - a_p \tag{28}$$

is primitive modulo $m$ [10, 12]. Again a sequence with maximum period qualifies as pseudo noise sequence since any $w$-tuple with $w \leq p$ appears precisely $m^{p-w}$ times within one period, except the all zero tuple which appears $m^{p-w} - 1$ times. The case $p = 1$, also known as linear congruential generator (LCG), has been proposed by D.H. Lehmer in 1949 [14]. With $m$ being limited by the wordsize of a computer, the period $m - 1$ is way too short for simulations that run on present day hardware. Yet the Lehmer generator has an interesting property: used as a pseudo random coin it yields $P_0(w) = 1/2$ for all values of $w$!

The LCG generates numbers $x_i \in \{1, 2, \ldots m-1\}$ and a natural way to simulate the flip of a coin is to choose "head" if $x_i \leq (m-1)/2$ and "tail" otherwise, i.e. to consider the binary sequence

$$y_k = \begin{cases} 0 & x_k \leq (m-1)/2 \\ 1 & \text{else} \end{cases}. \tag{29}$$

The period $T = m - 1$ of a LCG is even, and the pseudo noise property of the sequence $(x_k)$ guarantees that the sequence $(y_k)$ contains precisely $T/2$ ones and $T/2$ zeros. Setting $b = a_1^{T/2} \bmod m$ we have

$$\begin{aligned} x_{k+T/2} &= b x_k \bmod m \qquad \text{and} \\ x_{k+T} &= b^2 x_k = x_k \bmod m \,. \end{aligned} \tag{30}$$

Evidently $b = m - 1$ and $x_{k+T/2} \leq (m-1)/2$ if and only if $x_k > (m-1)/2$. For our binary sequence this means

$$y_{i+T/2} = y_i + 1 \bmod 2 \,, \tag{31}$$

i.e. every tuple is accompanied by its complementary tuple with zeros and ones exchanged. Eq. 31 immediately implies $P_0(w) = 1/2$.

Now let $(x_k)$ be a LFSR pseudo noise sequence with $p > 1$, and let $(x'_k)$ denote the sequence of every $d$-th element of $(x_k)$ i.e.

$$x'_k = x_{k_0+kd} \tag{32}$$

for fixed $k_0 < d$. $(x'_k)$ is called the $d$-decimated sequence of $(x_k)$. It can be proven that the decimation of a LFSR sequence yields again a LFSR sequence [6, 10]. For the particular choice

$$d = \frac{m^p - 1}{m - 1} = m^{p-1} + m^{p-2} + \cdots + 1 \tag{33}$$

the decimated sequence $(x'_k)$ is a LCG sequence, i.e. $x'_k = ax'_{k-1} \bmod m$ for some value $a$. This immediately implies that any zero in the original sequence is accompanied by zeros $d$ elements earlier and later,

$$x_k = 0 \quad \Longleftrightarrow \quad x_{k+d} = 0. \tag{34}$$

Apparently the distribution of zeros in LFSR pseudo noise sequences is not very random. For $m = 2$, $d$ equals the period and $x_k = x'_k$. For $m > 2$ we can apply the argument from above to the sequence $(x'_k)$ to get

$$x_{k+T/2} = (m-1)x_k \bmod m \tag{35}$$

for the original sequence with $T = m^p - 1$. All values in our sequence except zero are balanced in the sense that a small value $x_k$ is accompanied by a large value $m - x_k$ and vice versa. The bit sequence

$$y_k = \begin{cases} 0 & \text{if } 0 < x_k \leq (m-1)/2 \\ 1 & \text{if } (m-1)/2 < x_k < m \\ \text{ignore} & \text{if } x_k = 0 \end{cases} . \tag{36}$$

has $P_0(w) = 1/2$ for all $w$. Since only a fraction $1/m$ of all numbers are zeros, one does not have to ignore them explicitly if $m$ is large. In practical random number generators $m$ is close to the word size of the computer, like $m = 2^{31} - 1$, and the imbalance induced by the zeros can be neglected. For $m = 2$ it is dominant, however. This is the explanation why LFSR sequences over $\mathbb{Z}_2$ often fail in random walk simulations, whereas the same sequences on larger fields $\mathbb{Z}_m$, $m \gg 2$ work fine.

Strictly speaking we have proven $P_0(w) = 1/2$ for the sequence (36) only with respect to the complete period. It is a priori not clear whether simulations that explore a tiny fraction of the period can take advantage from the omission of zeros, so let us check this by a simple simulation. We consider the LFSR sequences

$$x_k = x_{k-32} + x_{k-63} \bmod 2 \tag{37}$$

and

$$x_k = x_{k-27} + x_{k-40} \bmod 3\,. \tag{38}$$

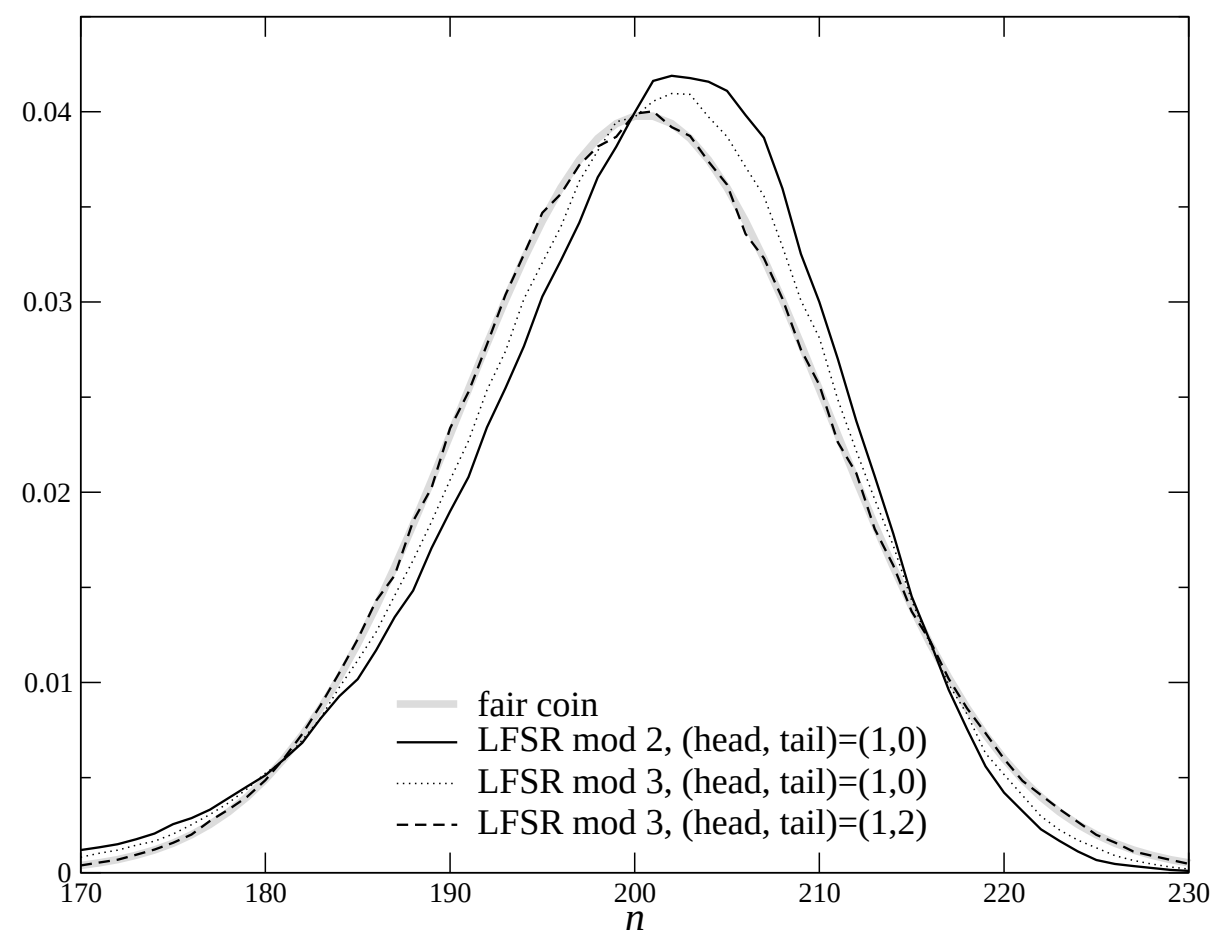

FIG. 11: Distribution of the number $n$ of "heads" in runs of length 401. The LFSR sequences are given by (37) and (38), the latter with two strategies to map $\{0, 1, 2\}$ to $\{$"head", "tail"$\}$. The data shown are averages over 512 000 non-overlapping tuples, a simulation that consumes less than $10^{-10}$ of the period of the underlying sequences.

Both sequences are pseudo noise sequences with comparable, long periods, $2^{63} - 1 \approx 9.2 \cdot 10^{18}$ and $3^{40} - 1 \approx 1.2 \cdot 10^{19}$. The binary sequence (37) is readily mapped onto $\{$"head", "tail"$\}$, but for the ternary sequence (38) we have the choice of which value to ignore. According to our preceding considerations it should make a difference whether the ignored value is zero or not.

Fig. 11 shows that this is indeed the case. We measured the distribution of "heads" in tuples of size 401. The sequence (37) has a significant bias towards "heads" if ones are interpreted as "heads" and zeroes as "tails". Almost the same bias can be observed in sequences from (38) if the same mapping between "heads" and "tails" and ones and zeroes is applied (and the value two is ignored). When the zeros are ignored and ones are interpreted as "heads" and twos as "tails", on the other hand, (38) is in perfect agreement with a fair coin. Note that the data shown are averages over 512 000 non-overlapping tuples. The total simulation consumes less than $10^{-10}$ of the period of the sequences. Apparently the zeros leave their traces in tiny samples.

The particular role of the zero is not surprising, considered its function in the arithmetic of fields: zero is the neutral element of the additive group and the only element that is not element of the multiplicative group. In LFSR sequences this shows in the reduced number of all zero tuples in a period and in the fact that the all zero tuple of size $p$ is the only fix point of the recursion. Another way to demonstrate the asymmetry of zero and non-zero values is to note that changing ones to zeros and vice versa in the initial seed of a LFSR sequence over $\mathbb{Z}_2$ does not result in the complementary sequence being generated [19]. What is surprising is *to what extent* the special nature of the zero affects random number generators in small fields like $\mathbb{Z}_2$ or $\mathbb{Z}_3$.

Note that we do not recommend linear recursions modulo 3 as pseudo random number generators. We have chosen (38) only to illustrate the special role of the zero. In our coin experiment we had to throw away $1/3$ of its output to get good results. Even for this simple simulation it is advisable to choose a large prime modulus $m$ to get a small fraction $1/m$ of zeros in the output stream. As a general rule, the modulus should be large in order to preserve the entropy of the pseudo random numbers under all circumstances [16].

## 6. THE RANDOM WALKER REVISITED

The original motivation of this work was to understand why recurrences in $\mathbb{Z}_2$ yield bad results when used as pseudo random number generators in random walk simulations. For one-dimensional random walks the relation between $P_0(w)$ and the position of the walker is obvious: $P_0(w)$ is the probability that after $w$ steps the walker ends to the left of his starting point (assuming that 0 encodes a step to the left).

For the walks in two dimensions discussed in [18], the interpretation depends on the specific algorithm used. The canonical way to generate a random walk is this:

```
(x, y) := (0, 0);
for t := 1 to w do
  if rand(0,1) = 1 then
    x := x + 1;
  else
    x := x − 1;
  if rand(0,1) = 1 then
    y := y + 1;
  else
    y := y − 1;
```

Here one movement along the diagonals of the square lattice is counted as a step, and each step consumes two pseudo random bits. The movement along the $x$- and $y$-direction is driven by the 2-decimated sequence of the original bit sequence, but decimating a LFSR sequence over $\mathbb{Z}_2$ by a power of 2 is equivalent to shifting the sequence [6, 19]. Vattulainen *et al.* generated $N$ random walks of length $w$ and counts how often the walker ends in each of the four quadrants of the lattice. They compare these values to the expected value $N/4$ using a $\chi^2$-test with three degrees of freedom. Equipped with our $P_0(w)$, we can calculate the $\chi^2$-value that is to be expected from this experiment. For simplicity we assume $w$ to be odd. Then the probability to end up in the southwest quadrant $(x, y < 0)$ reads $P_0^2(w)$, the probabilities for the other quadrants are $P_0(w)(1-P_0(w))$ (southeast and northwest) and $(1 - P_0(w))^2$ (northeast). The $\chi^2$-value of this distribution with respect to the uniform distribution reads

$$\frac{\chi^2}{N} = 3-16P_0(w)+32P_0^2(w)-32P_0^3(w)+16P_0^4(w)\,. \quad (39)$$

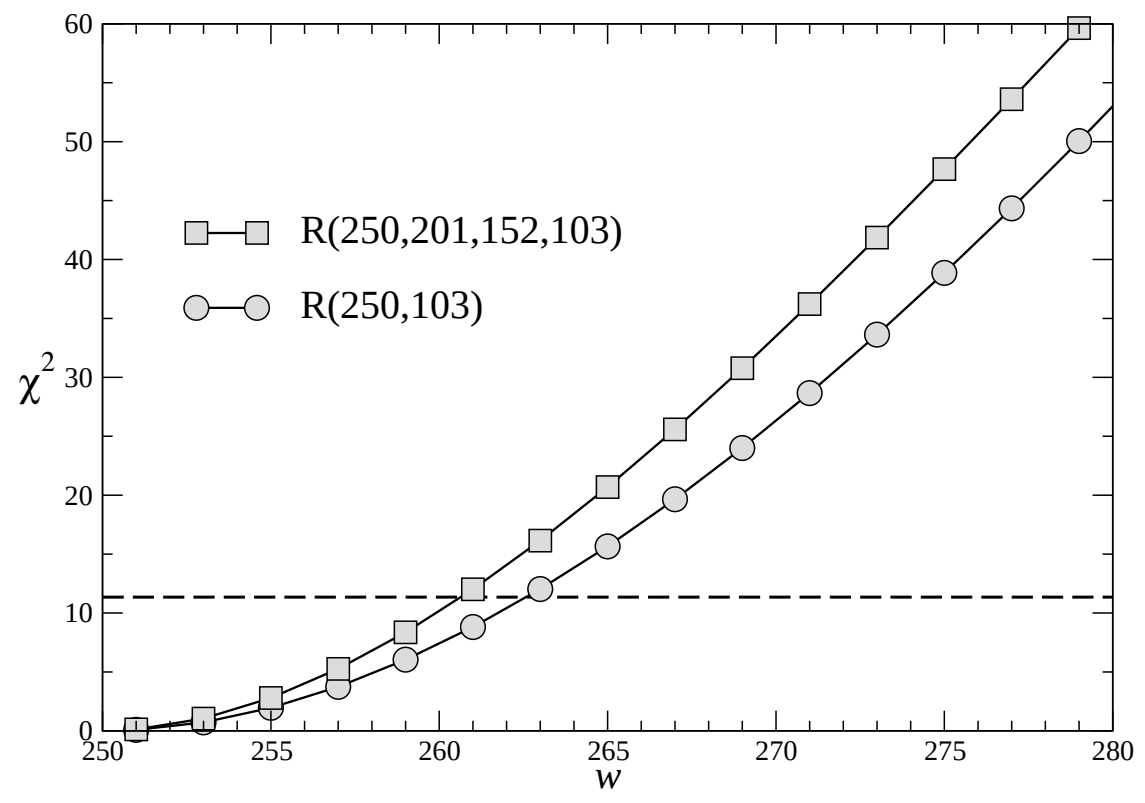

FIG. 12: $\chi^2$-value of the quadrant distribution in two-dimensional random walks of length $w$ as predicted by (39). The pseudo random bits are from R(250, 103) with $N = 10^6$ and from R(250, 201, 152, 103) with $N = 10^{10}$. The latter is equivalent to the 3-decimated R(250, 103). The dashed line indicates the 99 %-value of the $\chi^2$ distribution.

Fig. 12 shows $\chi^2$ for $N = 10^6$ and $P_0(w)$ from the R(250, 103) generator (Fig. 7). This corresponds to the experimental setup of Vattulainen *et al.* [18, Fig. 1]. The $\chi^2$ values increase as soon as the length $w$ of the walk gets larger than the size $p = 250$ of the shift register, the 99 %-line of the $\chi^2$-distribution being crossed at $w = 263$. The failure of R(250, 103) in random walk experiments is obvious.

Vattulainen *et al.* report that while the R(250, 103) fails the random walk test, its 3-decimated variant passes it. Decimating R(250, 103) by 3 results in the four tap generator R(250, 201, 152, 103) [19]. We can use Eq. (24) to calculate $P_0(w)$ and the corresponding value $\chi^2/N$ for $w$ not too far above 250. Fig. 12 shows that R(250, 201, 152, 103) fails the random walk test for samples of size $N \approx 10^{10}$. This holds of course for any other four tap generator with $p = 250$. Note that instead of increasing the number of samples one might as well increase the length of the walk to detect deviations, since in general $P_0(w)$ departs more and more from $1/2$ as $w$ increases, at least for quite a while.

The generalization of this analysis to walks in higher dimensions and to the $n$-block test discussed in [18] is straightforward.

Ziff [19] discusses extended simulations of kinetic self avoiding trails in two dimensions. Here a walker starts at the lower left hand corner of a $L \times L$ square lattice and heads in diagonal direction of the opposite corner. At each newly visited site the walker turns by 90 degrees either clockwise or anticlockwise with probability $1/2$ each.

When the walker encounters a site it has been visited before he always turns so as not to retrace its path. The lower and the left hand boundary of the lattice are reflective, while the upper and the right hand side are adsorbing. What is measured is the probability $P(L)$ that the walker is adsorbed by the upper boundary. Due to the symmetry $P(L)$ should equal $1/2$ for all values of $L$, but Ziff observed striking deviations in his simulations. Although there is no simple mapping between $P_0(w)$ and $P(L)$ knowledge of the former helps to understand the latter at least qualitatively. For example, the fact that two tap generators yield $P(L) < 1/2$ whereas $P(L) > 1/2$ for four tap generators corresponds nicely with the behavior of $P_0(w)$ (see Fig. 9 and Eq. (25) and (26)). Also the fact that $|P(L) - 1/2|$ increases with increasing $L$ with a rate that is smaller for four tap generators than it is for two tap generators is very similar to the behavior of $P_0(w)$. Eq. (24) allows us to calculate the sample sizes at which the generators $\mathrm{R}(9689, 471)$ and $\mathrm{R}(9689, 6988, 1586, 471)$ discussed by Ziff would fail the two dimensional random walk test of Vattulainen *et al.* The result is $10^8$ for the two tap generator and $10^{15}$ for the four tap generator. The latter value is out of reach for todays' computing power, but not for tomorrows'.

The simple random walks of Vattulainen *et al.* [18], Ziff's kinetic self avoiding trails [19] and the $3d$ self avoiding random walks discussed by Grassberger [7] all share a common feature: they show inconsistent results for pseudo random number generators that operate in $\mathbb{Z}_2$ and consistent results for generators that operate in $\mathbb{Z}_m$ with $m \approx 2^{31}$. Apparently the non-random behavior of the zeros in linear recurrences affects more complex simulations, too.

## 7. CONCLUSIONS

The main conclusion to be drawn from this paper is an advice: Do not produce pseudo random numbers using arithmetic in $\mathbb{Z}_2$, use arithmetic in $\mathbb{Z}_m$ instead, with $m$ being a large prime. Addition in $\mathbb{Z}_2$ is equivalent to the exclusive-or operation, an operation that is very very fast even when it is invoked from high-level languages like C, see [19] for a nice single line implementation. Addition in $\mathbb{Z}_m$ with $m$ being a prime $> 2$ requires a time consuming modulo-operation, hence it is comprehensible why some pseudo random number generators still operate in $\mathbb{Z}_2$. Our results in fact confirm the empirical recipes of improving the quality of modulo 2 generators, yet you should keep in mind that the clustering of zeros is suppressed but not truly removed by these measures. Pseudo random numbers generated by linear recurrences in $\mathbb{Z}_m$ have some deficiencies for large values of $m$, too. Random points in $d$-dimensional space are concentrated in hyperplanes for $d > p$ [13, 15], reflecting the linearity of the generating process. And there are long range correlations like the one shown in Eq. (35). None of these deficiencies seems to collide with practical applications, however. We are not aware of any simulation where LFSR sequences modulo 2 yield better results than LFSR sequences with modulo $m$ ($m$ large prime). With random walk simulations and cluster Monte Carlo simulations [16] we know at least two experiments where modulo $m$ generators are significantly better than modulo 2 generators.

Another conclusion of this work is that the empirically observed failure of some LFSR random number generators in random walk experiments can be explained theoretically. This actually argues in favor of the LFSR method, since it is better to know and to control the deficiencies of a random number generator than to rely on fancy methods which are basically justified by empirical observations. The LFSR method may appear old and outmoded, yet it fits perfectly with Donald Knuth's advice [12], "...random numbers should never be produced by a random method. Some theory should be used."

### Acknowledgments

It is a pleasure to thank Brian Hayes for stimulating discussions. This work was supported by the German science council (Deutsche Forschungsgemeinschaft) under grant ME2044/1-1.